\documentstyle[12pt,graphicx,epsf]{article}
\begin{document}

\def\AEF{A.E. Faraggi}

\def\vol#1#2#3{{\bf {#1}} ({#2}) {#3}}
\def\NPB#1#2#3{{\it Nucl.\ Phys.}\/ {\bf B#1} (#2) #3}
\def\PLB#1#2#3{{\it Phys.\ Lett.}\/ {\bf B#1} (#2) #3}
\def\PLA#1#2#3{{\it Phys.\ Lett.}\/ {\bf A#1} (#2) #3}
\def\PRD#1#2#3{{\it Phys.\ Rev.}\/ {\bf D#1} (#2) #3}
\def\PRL#1#2#3{{\it Phys.\ Rev.\ Lett.}\/ {\bf #1} (#2) #3}
\def\PRT#1#2#3{{\it Phys.\ Rep.}\/ {\bf#1} (#2) #3}
\def\MODA#1#2#3{{\it Mod.\ Phys.\ Lett.}\/ {\bf A#1} (#2) #3}
\def\RMP#1#2#3{{\it Rev.\ Mod.\ Phys.}\/ {\bf #1} (#2) #3}
\def\IJMP#1#2#3{{\it Int.\ J.\ Mod.\ Phys.}\/ {\bf A#1} (#2) #3}
\def\nuvc#1#2#3{{\it Nuovo Cimento}\/ {\bf #1A} (#2) #3}
\def\RPP#1#2#3{{\it Rept.\ Prog.\ Phys.}\/ {\bf #1} (#2) #3}
\def\APJ#1#2#3{{\it Astrophys.\ J.}\/ {\bf #1} (#2) #3}
\def\APP#1#2#3{{\it Astropart.\ Phys.}\/ {\bf #1} (#2) #3}
\def\EJP#1#2#3{{\it Eur.\ Phys.\ Jour.}\/ {\bf C#1} (#2) #3}
\def\JHP#1#2#3{{\it JHEP}\/ {\bf #1} (#2) #3} 
\def\etal{{\it et al\/}}

\newcommand{\cc}[2]{c{#1\atopwithdelims[]#2}}
\newcommand{\bev}{\begin{verbatim}}
\newcommand{\beq}{\begin{equation}}
\newcommand{\beqa}{\begin{eqnarray}}
\newcommand{\beqn}{\begin{eqnarray}}
\newcommand{\eeqn}{\end{eqnarray}}
\newcommand{\eeqa}{\end{eqnarray}}
\newcommand{\eeq}{\end{equation}}
\newcommand{\beqt}{\begin{equation*}}
\newcommand{\eeqt}{\end{equation*}}
\newcommand{\Eev}{\end{verbatim}}
\newcommand{\bec}{\begin{center}}
\newcommand{\eec}{\end{center}}
\def\ie{{\it i.e.}}
\def\eg{{\it e.g.}}
\def\half{{\textstyle{1\over 2}}}
\def\nicefrac#1#2{\hbox{${#1\over #2}$}}
\def\third{{\textstyle {1\over3}}}
\def\quarter{{\textstyle {1\over4}}}
\def\m{{\tt -}}
\def\mass{M_{l^+ l^-}}
\def\p{{\tt +}}

\def\slash#1{#1\hskip-6pt/\hskip6pt}
\def\slk{\slash{k}}
\def\GeV{\,{\rm GeV}}
\def\TeV{\,{\rm TeV}}
\def\y{\,{\rm y}}

\def\l{\langle}
\def\r{\rangle}
\def\LRS{LRS  }

\begin{titlepage}
\samepage{
\setcounter{page}{1}
\rightline{LTH--943} 
\vspace{1.5cm}
\begin{center}
 {\Large \bf The Equivalence Postulate of Quantum Mechanics,\\ 
             Dark Energy and \\ 
             The Intrinsic Curvature of Elementary Particles }
\vspace{.25 cm}

Alon E. Faraggi\footnote{
		                  E-mail address: alon.faraggi@liv.ac.uk}
\\
\vspace{.25cm}
{\it Department of Mathematical Sciences\\
University of Liverpool, Liverpool, L69 7ZL, United Kingdom}
\end{center}

\begin{abstract}

The equivalence postulate of quantum mechanics offers an axiomatic
approach to quantum field theories and quantum gravity. The equivalence
hypothesis can be viewed as adaptation of the classical Hamilton--Jacobi
formalism to quantum mechanics. The construction reveals two
key identities that underly the formalism in Euclidean or Minkowski spaces.
The first is a cocycle condition, which is invariant under $D$--dimensional 
Mobi\"us transformations with Euclidean or Minkowski metrics. 
The second is a quadratic identity which is a representation of the 
$D$--dimensional quantum Hamilton--Jacobi equation.
In this approach, the solutions of the associated Schr\"odinger equation are
used  to solve the non--linear quantum Hamilton--Jacobi equation. 
A basic property of the construction is that the two solutions
of the corresponding Schr\"odinger equation must be retained.
The quantum potential, which arises in the formalism, can be interpreted
as a curvature term.  I propose that the quantum potential, which is always
non--trivial and is an intrinsic energy term characterising a particle, can 
be interpreted as dark energy. Numerical estimates of its magnitude 
show that it is extremely suppressed. In the multi--particle case the 
quantum potential, as well as the mass, are cumulative.

\end{abstract}
\smallskip}
\end{titlepage}

\section{Introduction}

Understanding the synthesis of quantum mechanics and gravity is 
an important challenge in theoretical physics.
The main effort in this endeavour is in the framework of string theory. 
The primary advantage of string theory is that it gives rise
to the gauge and matter ingredients of elementary particle 
physics, and predicts the number of degrees of freedom needed to 
obtain a consistent theory. 
String theory therefore enables the construction
of quasi--realistic models and the development 
of a phenomenological approach to quantum gravity.
The state of the art in this regard are the heterotic string models 
in the free fermionic formulation, which reproduce the matter
content of the Minimal Supersymmetric Standard Model and 
preserve its unification picture \cite{ffslm}. 
Despite its phenomenological success string theory does not provide a
framework for a rigorous formulation of quantum gravity from 
fundamental principles. 

Important characteristics of string theories are its various perturbative and 
non--perturbative dualities. Indeed, one of the interesting approaches to
formulating string theory aims to promote $T$--duality \cite{gpr}
to a manifest property of the formalism \cite{hull}. 
$T$--duality can be viewed heuristically as phase--space duality in compact space. 

A formalism that aims to promote phase--space duality to a level of 
a fundamental principle was followed in the context of the 
equivalence postulate approach to quantum mechanics \cite{fm,bfm}. 
The equivalence postulate of quantum mechanics
hypothesizes that all physical systems are equivalent
under coordinate transformations. In particular, there should always 
exist a coordinate transformation connecting a physical system with
a non--trivial potential $V$ and energy $E$, to the one with 
$V-E=0$. Conversely, any allowed physical state should arise
by a coordinate transformation from the state with $V-E=0$.
Thus, non--trivial states arise from the inhomogeneous term
stemming from the transformation of the trivial state
under coordinate transformations. It is then seen 
that both the definability of the phase--space duality, 
as well as consistency of the equivalence postulate for
all physical states requires the modification of classical mechanics
by quantum mechanics. More precisely, the phase--space duality, 
as well as the equivalence postulate, are ill defined in classical 
mechanics for the trivial 
state, for which Hamilton's generating function $S_0$ is a constant
or a linear function of the coordinate. The quantum 
modification removes this state from the space of allowed solutions, 
and enables the consistency of the equivalence postulate, 
as well as definability of phase--space duality, for all physical states. 

In this paper, I argue that another feature of the equivalence postulate 
approach is the existence of dark energy, which arises as an intrinsic 
property of elementary particles.
This property of elementary particles arises from 
the quantum potential, which is never vanishing
and correspond to an intrinsic curvature associated 
with elementary particles. 

\section{Equivalence postulate of quantum mechanics} 

In this section I review the equivalence postulate 
approach to quantum mechanics. It is important to 
emphasize that the equivalence postulate formulation
of quantum mechanics does not entail an interpretation
or a modification, but rather 
a mathematically rigorous derivation of quantum mechanics 
from a fundamental postulate. In this respect the equivalence 
postulate provides an axiomatic approach
to quantum mechanics and quantum field theories.
The formalism has some reminiscences with Bohm's approach
to quantum mechanics in the sense that both approaches
may be regarded as a modification of the classical Hamilton--Jacobi 
formalism. However, aside from this superficial similarity
the two approaches are fundamentally distinct.

A view of the equivalence postulate approach is obtained by an
analogy with the classical Hamilton--Jacobi formalism. 
The classical Hamilton equations of motion are 
given by
\beq
{\dot p}= -{{\partial H}\over {\partial q}} ~~{\rm and}~~
{\dot q}= {{\partial H}\over {\partial p}}
\label{Heqsofmotion}
\eeq
where $p$ and $q$ are the phase space variables, and $H$ is
the Hamiltonian. In classical 
mechanics the solution to the mechanical problem is obtained
by performing canonical transformations to 
a new set of phase space variables
\beq
q\rightarrow Q(q,p,t) , p\rightarrow P(q,p,t)
\label{canonicaltrans}
\eeq
such that the new Hamiltonian and Hamilton equations of 
motion are trivial $H(q,p) \rightarrow K(Q,P)\equiv 0$.
Consequently, the transformed phase space coordinates 
are constants of the motion. 
The solution to this problem in classical mechanics is
given by the Classical Hamilton--Jacobi Equation (CHJE) 
which for systems that conserve the total energy gives 
the Classical Stationary Hamilton--Jacobi Equation (CSHJE), 
\beq
{1\over{2m}}\left({{\partial S}\over {\partial q}}\right)^2+ V(q) = E
\label{cshje}
\eeq
where $S(q)$ is a generating function.
For the purpose of this paper it is sufficient to focus here
on the stationary case. 
In performing these trivialising transformations 
the phase space variables are taken as independent variables. 
The functional dependence between them is only extracted 
after solving the CSHJE eq. (\ref{cshje}) from the 
relation 
\beq
p={{\partial S(q)}\over {\partial q}}.
\label{peqpsdq}
\eeq
Since the transformations (\ref{canonicaltrans}) are invertible 
we also have the inverse transformation 
\beq
Q\rightarrow q(Q,P,t) , P\rightarrow p(Q,P,t). 
\label{invtrans}
\eeq
However, the intrinsic property of quantum mechanics is that 
the phase space variables are not independent, {\it i.e.} they 
satisfy the commutation relation, 
\beq
\left[{\hat q},{\hat p}\right] = i\hbar. 
\label{qmcommutationrelation}
\eeq
We therefore relax the condition that the phase space variable are independent
in the application of the trivialising transformation, and hence that
the transformations are canonical. We further assume the functional 
relation between the phase--space variables via the generating function, 
eq. (\ref{peqpsdq}). 
we can ask the following question: is it possible in classical 
mechanics to start with a system with a non--trivial 
Hamiltonian, {\it i.e.} with a given nonvanishing $W(q)$, 
and connect via some 
coordinate transformations to any other system.
This demand dictates that the Hamilton--Jacobi equation 
retains its form under the coordinate transformations. 
From the form of eq. (\ref{cshje}) we see that this is not possible 
in classical mechanics. The reason is that in classical 
mechanics the trivial state, with $V(q)-E\equiv0$,  
is a fixed state under the transformations. Insisting that 
all physical states, including the trivial one, are connected 
by coordinate transformations necessitates the modification of the
CSHJE. The modification is given by the Quantum Stationary 
Hamilton--Jacobi Equation (QSHJE)
\beq
{1\over{2m}}\left({{\partial S(q)}\over {\partial q}}\right)^2+ W(q)+Q(q)=0~,
\label{qshje}
\eeq
where $W(q)=V(q)-E$. From the form of eq. (\ref{qshje}) 
it is seen that the combination $W(q)+Q(q)$ transforms 
as a quadratic differential under coordinate transformations, 
whereas each of the functions $W(q)$ and $Q(q)$ transform
as a quadratic differential up to an additive term, {\it i.e.}
under $q\rightarrow {\tilde q}(q)$ we have, 
\beqn
W(q)\rightarrow  {\tilde W}({\tilde q}) & = &
              \left( {{\partial q}\over {\partial{\tilde q}}}\right)^2
                        {W}({q}) + (q;{\tilde q})\nonumber\\
Q(q)\rightarrow  {\tilde Q}({\tilde q}) & = &
              \left( {{\partial q}\over {\partial{\tilde q}}}\right)^2
                        {Q}({q}) - (q;{\tilde q}).\nonumber
\eeqn
and 
\beq
(W(q)+Q(q))\rightarrow  ({\tilde W}({\tilde q})+
                         {\tilde Q}({\tilde q}))= 
              \left( {{\partial q}\over {\partial{\tilde q}}}\right)^2
                         (W(q)+Q(q))\nonumber
\eeq 
All physical states with a non--trivial $W(q)$ then arise 
from the inhomogeneous part in the transformation of the 
trivial state $W^0(q^0)\equiv0$, {\it i.e.} $W(q)= (q^0; q)$. 

Considering the transformation $q^a\rightarrow q^b\rightarrow q^c$
versus $q^a\rightarrow q^c$ gives rise to the cocycle
condition on the inhomogeneous term 
\beq
(q^a;q^c)= \left({{\partial q^b}\over {\partial q^c}}\right)^2
           \left[ (q^a;q^b) - (q^c;q^b)\right]. 
\label{cocycle}
\eeq
The cocycle condition, eq. (\ref{cocycle}) underlies the equivalence
postulate, and embodies its underlying symmetries. In particular, it
is invariant under the Mobius transformations, 
\beq
(\gamma(q^a);q^b)=(q^a;q^b),
\label{am21}\eeq
where
\beq
\gamma(q)={{Aq+B}\over {Cq+D}}
\label{mobiustrans}
\eeq
and 
$\left(\begin{array}{c}A\\C\end{array}\begin{array}{cc}B\\D\end{array}
\right)\in GL(2,{C}).$ In one dimension the cocycle condition
(\ref{cocycle}) uniquely defines the Schwarzian derivative
up to a constant and a coboundary term. Specifically, one obtains
$(q^a; q^b)=-\beta \{ q^a, q^b\}/{4m}$, where
$\{f,q\} =f^{\prime\prime\prime}/f^\prime-3(f^{\prime\prime}/f^\prime)^2/2$ denotes
the Schwarzian derivative and $\beta$ is a constant
with the dimension of an action. We further have that 
$\{\gamma(q), q\}\equiv 0$.   

The one dimensional stationary case is instructive to 
reveal the symmetry properties that underlie quantum 
mechanics in the equivalence postulate formalism. In one dimension 
the unique solution of the problem is given
in terms of the Schwarzian identity
\beq
\left({{\partial S(q)}\over {\partial q}}\right)^2=
{\beta^2\over 2}
\left(\left\{{\rm e}^{{{2i}\over\beta}{S}},q\right\}-\left\{S,q\right\}\right)
\label{schwarzianidentity}
\eeq
which embodies the equivalence postulate, and leads to the Schr\"odinger
equation. Making the identification
\beq
W(q)= V(q)-E = -{\beta^2\over {4m}}\left\{{\rm e}^{{{(i2S_0)}\over \beta}},q\right\},
\label{wqeqvqminuse}
\eeq
and 
\beq
Q(q) = {\beta^2\over {4m}}\left\{S_0,q\right\}, 
\label{qq}
\eeq
we have that $S_0$ is the solution of the Quantum Stationary Hamilton--Jacobi
equation (QSHJE), 
\beq
{1\over {2m}}\left({{\partial S_0}\over {\partial q}}\right)^2 + V(q) - E + {\hbar^2\over{4m}}
\left\{S_0,q\right\} = 0 , 
\label{qshje2}
\eeq
From, eq. (\ref{wqeqvqminuse}), and
the properties of the Schwarzian derivative, we deduce that $S_0$, 
the solution of the QSHJE eq. (\ref{qshje2}) is given by (see also \cite{floyd}),
\beq
{\rm e}^{{2i\over \beta}S_0}= \gamma(w)= {{Aw+B}\over {Cw +D}} = {\rm e}^{i\alpha}
{{w+i{\bar\ell}}\over {w-i\ell}}
\label{ei2s0overbeta}
\eeq
where $\ell= \ell_1+i\ell_2$; $\{\alpha, \ell_1, \ell_2\}\in R$.  
Here $w=\psi^D/\psi$ and $\psi^D$ and $\psi$ are two linearly independent solutions
of a second order differential equation given by
\beq
\left(-{\beta^2\over {2m}} {\partial^2\over {\partial q}^2} + V(q) - E \right)\psi(q)=0
\label{se}
\eeq
{\it i.e.} $\psi^D$ and $\psi$ are the two solutions of the Schr\"odinger equation
and we can identify $\beta\equiv\hbar$. We can note the relation between the 
Shr\"odinger equation in an alternative way. Inserting the solution
\beq
\psi= R{\rm e}^{{i\over\hbar}S_0}
\label{psiq}
\eeq 
into the Schr\"odinger equation produces the two equations
\beqn
&\left(\partial_q S_0\right)^2+ V(q)-E-
\hbar^2 (\partial_q^2 R)/(2mR)=0 ~, & 
\label{qhjefse}\\
&\partial_q \left(R^2\partial_q S_0\right)= 0~.
\label{ce}
\eeqn
The continuity equation, eq. (\ref{ce}), gives $R=1/\sqrt{S_0^\prime}$ and 
consequently 
\beq
Q(q) = -{\hbar^2\over {2m}}{{\partial_q^2 R}\over R}= {\hbar^2\over {4m}}
\{S_0,q\} 
\label{qqReqS0p}
\eeq
and eq. (\ref{qhjefse}) corresponds to eq. (\ref{qshje2}). 

The steps taken in deriving the quantum Hamilton--Jacobi equation
eq. (\ref{qshje2}) from the Schr\"odinger equation are reminiscent 
of its derivation in the framework of Bohmian quantum mechanics. However, 
there is a crucial difference. While in Bohmian mechanics 
one identifies the solution (\ref{psiq}) with the wave function, 
and hence $R^2$ with the probability density, it is noted that
the equivalence postulate necessitates that both 
solutions $\psi$ and $\psi^D$ are kept in the formalism. This can be seen
from the properties of the Schwarzian derivative that show that
the trivialising transformation is given by
\beq
q\rightarrow q^0 \equiv \gamma(\psi^D/\psi)
\label{qtoq0}
\eeq
{\it i.e.} up to a Mobius transformation, $q^0$ is given by the ratio
of the two solutions of the corresponding Schr\"odinger equation. 
Hence, in general the wave function in the equivalence postulate 
approach is given by 
\beq
\Psi(q) = R(q)\left(A {\rm e}^{{i\over\hbar}S_0} + B {\rm e}^{-{i\over\hbar}S_0}\right)
\label{Psiq}
\eeq
Furthermore, the equivalence postulate necessitates that
$S_0(q)\ne Aq +B$, {\it i.e.} $S_0(q)$ cannot be a linear function of $q$. 
Strictly speaking the condition that the Schwarzian derivative 
is well defined only necessitates the weaker condition $S_0(q)\ne constant$. 
However, the condition that the quantum potential is always non--vanishing
leads to the stronger constraint $S_0(q)\ne Aq +B$, 
which also follows from arguments concerning phase--space duality
\cite{fm}. 
In the time dependent case the equivalence postulate implies 
that the wave function should always take the form
\beq
\Psi(q,t) = R(q,t)\left(A {\rm e}^{{i\over\hbar}S(q,t)} + B {\rm e}^{-{i\over\hbar}S(q,t)}\right), 
\label{Psiq2}
\eeq
where $S=S_0-Et$ in the stationary case. 
As discussed above, consistency of the equivalence postulate of
quantum mechanics dictates the necessity of employing the two
solutions of the Schr\"odinger equation. 
This condition is well known in relativistic quantum mechanics
and signals the departure from the single particle interpretation
in non--relativistic quantum mechanics to the multi particle
representation  of quantum field theories. 
However, in non--relativistic Bohmian mechanics, 
and in particular in the case of bound states,
only the solution with the positive exponent is kept, 
which implies that in those cases in Bohmian mechanics
$S_0=constant$. 
 The equivalence postulate approach, 
on the other hand necessitates that both solutions
are kept in the formalism, and that $S_0\ne0$ always.
Consistency of the equivalence postulate further
implies that the trivialising map, 
$q\rightarrow {\tilde q} =\psi^D/\psi$, 
is continuous on the extended real line \cite{fm}. It is then seen that this
condition is synonymous with the requirement
that the physical solution of the corresponding 
Schr\"odinger equation admits a square integrable 
solution, and selects the correct physical eigenstates
for the bound states. In the relativistic case the inclusion of the
"negative energy" states reveals the existence of anti--particles. 
The implications of incorporating the 
two solutions in the non--relativistic case requires more detailed scrutiny. 
It is noted that this decomposition of the wave--function has been 
employed successfully in studies of molecular dynamics, \cite{poirier} and
is referred to there as the bi--polar decomposition. 

The equivalence postulate formalism extends to the higher dimensional case
both with respect to the Euclidean and Minkowski metrics \cite{bfm}. The key to 
these extensions are the generalisations of the cocycle condition eq. 
(\ref{cocycle}), and of the quadratic identity eq. (\ref{schwarzianidentity}). 
Denoting the transformations between two sets of coordinate systems 
by
\beq
q\rightarrow q^v = v(q)
\label{qtovq}
\eeq
and the conjugate momenta by the generating function $S_0(q)$,
\beq
p_k= {{\partial S_0}\over {\partial q_k}}. 
\label{pkmomenta}
\eeq
Under the transformations (\ref{qtovq}) we have $S_0^v(q^v)=S_0(q)$, hence
\beq
p_k\rightarrow p_k^v=\sum_{i=1}^D J_{ki}p_i
\label{pkv}
\eeq
where  $J$ is the Jacobian matrix
\beq
J_{ki}={{\partial q_i}\over {\partial q_j^v}}.
\label{jacobian}
\eeq
Introducing the notation
\beq
(p^v|p)={{\sum_k (p_k^v)^2}\over{\sum_kp_k^2}}={{p^tJ^tJp}\over {p^tp}}. 
\label{pvp}
\eeq
 the cocycle condition takes the form 
\beq
(q^a;q^c)=(p^c|p^b)\left[(q^a;q^b)-(q^c;q^b)\right], 
\label{cocycleinEspace}
\eeq
which captures the symmetries that underly quantum mechanics. 
It is shown that the cocycle condition, eq. (\ref{cocycleinEspace})
is invariant under $D$--dimensional Mobi\"us transformations, 
which include dilatations, rotations, translations and reflections in
the unit sphere \cite{bfm}. The quadratic identity, eq.  (\ref{schwarzianidentity}), 
is generalised by the basic identity
\beq 
\alpha^2(\nabla S_0)^2=
{\Delta(R{\rm e}^{\alpha S_0})\over R{\rm e}^{\alpha S_0}}-{\Delta R\over R}-{\alpha\over R^2}\nabla\cdot(R^2\nabla S_0), 
\label{ddidentity}
\eeq
which holds for any constant $\alpha$ and any functions $R$ and $S_0$. Then, if $R$ 
satisfies the continuity equation 
\beq
\nabla\cdot(R^2\nabla S_0)=0, 
\label{conteq}
\eeq
and setting $\alpha=i/\hbar$, we have 
\beq
{1\over2m}(\nabla S_0)^2=-{\hbar^2\over2m}{\Delta(R{\rm e}^{{i\over 
\hbar} S_0})\over R{\rm e}^{{i\over\hbar}S_0}}+
{\hbar^2\over2m}{\Delta R\over R}. 
\label{identity2}
\eeq 
In complete analogy with the one dimensional case we make identifications, 
\beqn
W(q)=V(q)-E& = &{\hbar^2\over2m}{\Delta(R{\rm e}^{{i\over\hbar}S_0})\over 
R{\rm e}^{{i\over \hbar}S_0}}, 
\label{ddwq}\\
Q(q)& =& -{\hbar^2\over2m}{\Delta R\over R}. 
\label{identity3}
\eeqn
Eq. (\ref{ddwq}) implies the $D$--dimensional Schr\"odinger equation
\beq
\left[-{\hbar^2\over2m}{\Delta}+V(q)\right]\Psi=E\Psi. 
\label{ddschroedingereq}
\eeq
and the general solution
\beq
\Psi= R(q) \left( A {\rm e}^{{i\over \hbar} S_0} + 
B {\rm e}^{-{i\over \hbar} S_0}\right).
\label{ddwavefunction}
\eeq
is mandated by consistency of the equivalence postulate. 

The equivalence postulate formalism generalises to the 
relativistic case as well. 
In this case, setting $q\equiv (ct, q_1, \ldots, q_{D-1})$, 
with $q^v=v(q)$ a general 
transformation of the coordinates, we have  
\beq
(p^v|p)={\eta^{\mu\rho}p_\mu^vp_\rho^v\over\eta^{\mu\rho}p_\mu p_\nu}=
{p^tJ\eta J^tp\over 
p^t\eta p}, 
\label{pvprelativistic}
\eeq
and $J$ is the Jacobian matrix 
\beq
{J^\mu}_\rho={\partial q^\mu\over\partial{q^v}^\rho}. 
\label{relativisticjacobian}
\eeq
Furthermore, we obtain the cocycle condition 
\beq
(q^a;q^c)=(p^c|p^b)\left[(q^a;q^b)-(q^c;q^b)\right], 
\label{relcocycle}
\eeq
and is invariant under $D$--dimensional Mobi\"us transformations with respect to
Minkowski metric. The quadratic identity in this case takes the form, 
\beq
\alpha^2(\partial S)^2={\Box(R{\rm e}^{\alpha S})\over R{\rm e}^{\alpha S}} 
-{\Box R\over R}-{\alpha\over R^2}\partial \cdot (R^2\partial S), 
\label{qidentityrel}
\eeq
which holds for any constant $\alpha$ and any functions $R,$ and $S$. 
Then, if $R$ satisfies 
the continuity equation $\partial(R^2\cdot\partial S)=0$, and setting 
$\alpha=i/\hbar$ we have 
\beq 
(\partial S)^2=-{\hbar^2}{\Box(R{\rm e}^{{i\over 
\hbar} S})\over R{\rm e}^{{i\over\hbar} S}}+{\hbar^2}{\Box R\over R}. 
\label{identity2rel}
\eeq
Setting 
\beqn
W(q) ~=~ {mc^2} & = & -{\hbar^2}{\Box(R{\rm e}^{{i\over 
\hbar} S})\over R{\rm e}^{{i\over\hbar} S}} \label{wqrelativistic}\\ 
Q(q) & =& {\hbar^2}{\Box R\over R} \label{qqrelativistic}
\eeqn
reproduces the relativistic Klein--Gordon equation
\beq
\left( {\hbar^2}{\Box}+ mc^2\right) \Psi(q) = 0 \label{rkge}
\eeq
with the general solution
\beq
\Psi  = R(q) ( A  {\rm e}^{{i\over  \hbar} S} + 
               B  {\rm e}^{-{i\over  \hbar} S}). 
\label{rwf}
\eeq
The equivalence postulate formalism also incorporates gauge interactions 
via the generalisation of the quadratic identities, eqs. 
(\ref{schwarzianidentity}, \ref{ddidentity}, \ref{qidentityrel}),
and of the cocycle 
conditions, eqs (\ref{cocycle}, \ref{cocycleinEspace}, \ref{relcocycle}).
In the nonrelativistic stationary case the quadratic identity takes the form 
\beq
\alpha^2(\nabla S_0 +eA)^2=
{(\nabla+e\alpha A)^2(R{\rm e}^{\alpha S_0})\over R{\rm e}^{\alpha S_0}}-
{\Delta R\over R}-{\alpha\over R^2}\nabla\cdot(R^2(\nabla S_0+eA)), 
\label{gaugepotID}
\eeq
and the cocycle condition eq. (\ref{cocycleinEspace}) retains its form with
\beq
(p^v|p)={{(p^v+eA^v)^t(p^v+eA^v)}\over {(p+eA)^t(p+eA)}}.
\label{pavpa}
\eeq
Whereas in the relativistic case they take the form
\beq
\alpha^2(\partial S+eA)^2={D^2Re^{\alpha S}\over Re^{\alpha S}} 
-{\Box R\over R}-{\alpha\over R^2}\partial\cdot(R^2(\partial S+eA)). 
\label{relgaugepotID}
\eeq
where 
\beq
D_{\mu}=\partial_{\mu}+\alpha eA_{\mu}, 
\label{ilcovariante}
\eeq
and the cocycle identity (\ref{cocycleinEspace}) retains its form with 
\beq
(p^b|p)={(p^b+eA^b)^2\over(p+eA)^2}={(p+eA)^tJ\eta J^t(p+eA) 
\over(p+eA)^t\eta(p+eA)}, 
\label{relcocyclewithpot}
\eeq 
and $J$ is the Jacobian matrix 
\beq
{J^\mu}_\nu={\partial q^\mu\over\partial{q^b}^\nu}. 
\label{opsiqj}
\eeq
We note the symmetry structure that underlies the formalism. 
Seeking further generalisation of this approach simply entails that 
this robust symmetry structure is retained. While the formalism 
has some reminiscences with Bohmian quantum mechanics, it is clearly
distinct from it as no physical interpretation has so far been assigned 
to the wave function. Furthermore, consistency of the equivalent
postulate formalism requires that the physical solutions are
square integrable and selects the same eigenstates that 
in conventional quantum mechanics arise due to the 
probability interpretation of the wave function. In the 
framework of the equivalence postulate, the Schr\"odinger
equation, and its solutions, is merely a device to find 
solutions of the non--linear quantum Hamilton--Jacobi 
equation. 
To investigate the
physical picture of the quantum potential in the
equivalence postulate formalism we have to turn
to its interpretation as a curvature term.

\section{The quantum potential as a curvature term}

The 
Quantum Stationary Hamilton--Jacobi Equation (QSHJE),
eq. (\ref{qshje})
can be viewed as a deformation of the
Classical Stationary Hamilton--Jacobi Equation (CSHJE), 
eq. (\ref{cshje}), 
by a ``conformal factor''. 
Noting that
\beq
\{S_0,q\}=-(\partial_q S_0)^2\{q, S_0\},
\label{inverseS0tran}
\eeq
we have that the QSHJE (\ref{qshje}) is equivalent to
\beq
{1\over2m}\left({\partial S_0\over\partial q}\right)^2
\left[1-\hbar^2 U( S_0)\right]+V(q)-E=0,
\label{qshje3}
\eeq
where $U(S_0)$ is the canonical potential
\beq
U(S_0)={1\over2}\{q, S_0\},
\label{pqdpotential}
\eeq
that arises in the framework of the $p$--$q$ duality \cite{fm}.
Eq.(\ref{qshje3}) can be written in the form
\beq
{1\over2m}\left({\partial S_0\over\partial\hat q}\right)^2+V(q)-E=0,
\label{qshje4}
\eeq
where
\beq
\left({\partial q\over\partial\hat q}\right)^2=\left[1-\hbar^2 U(S_0)\right],
\label{pqopqhat}
\eeq
or equivalently
\beq
d\hat q={dq\over\sqrt{1-\beta^2(q)}},
\label{pqopqhat2}
\eeq
with $\beta^2(q)=\hbar^2 U(S_0)=\hbar^2\{q, S_0\}/2$.
Integrating
(\ref{pqopqhat}) yields
\beq
\hat q=\int^q{dx\over\sqrt{1-\beta^2(x)}}.
\label{quantumcoordinate}
\eeq
It follows that
\beq
\lim_{\hbar\longrightarrow0}\hat q=q.
\label{limhbarqhat}
\eeq
In the case of the trivial state $W^0(q^0)\equiv0$, 
Eq.(\ref{qshje3}) becomes
\beq
{1\over2m}\left({\partial S_0^0\over\partial q^0}\right)^2\left[1-\hbar^2
 U(S_0)\right]=0,
\label{qshje5}
\eeq
whose solution is
\beq
U\left({\hbar\over2i}\ln\gamma(q^0)\right)={1\over\hbar^2},
\label{U0solution}
\eeq
where
\beq
\gamma(q^0)={Aq^0+B\over Cq^0+D},
\label{gammaqzero}
\eeq
with $AD-BC\ne 0$. 
Eq. (\ref{qshje3}) shows that the quantum potential can be regarded
as a deformation of the space--geometry. From eqs. 
(\ref{gaugepotID}, \ref{relgaugepotID}) we note that in this
respect the quantum potential is distinct from the 
interactions that arise from the gauge potentials, 
that can be viewed as a shift of the momentum.
The quantum potential has a universal character,
which is independent of gauge charges. Furthermore,
Flanders showed that
the Schwarzian derivative can be interpreted as a
curvature of an equivalence problem for curves in 
${\bf P}^1$ \cite{Flanders}. For that purpose,   
introduce a frame for
${\bf P}^1$, that is a pair ${\bf x},{\bf y}$ of points
in affine space ${\bf A}^2
$ such that $[{\bf x},{\bf y}]=1$, where
\beq
[{\bf x},{\bf y}]={\bf x}^t\left(\begin{array}{c} 0\\ -1
\end{array}\begin{array}{cc}1\\ 0\end{array}\right){\bf y}=x_1y_2-x_2y_1,
\label{affinespace}
\eeq
is the area function, which has the
$SL(2,{\bf R})$--symmetry
\beq
[\tilde{\bf x},\tilde{\bf y}]=[{\bf x},{\bf y}],
\label{areasymmetry}
\eeq
where $\tilde{\bf x}=R{\bf x}$ and $\tilde{\bf y}=R{\bf y}$ with 
$R\in SL(2, {\bf R})$.
Considering the moving frame $s\longrightarrow\{{\bf x}(s),{\bf y}(s)\}$ 
and differentiating $[{\bf x},{\bf y}]=1$ yields the structure equations
\beq
{\bf x}'=a{\bf x}+b{\bf y},\qquad{\bf y}'=c{\bf x}-a{\bf y},
\label{structureequations}
\eeq
where $a,b$, and $c$ depend on $s$. Given a map $\phi=\phi(s)$ from a domain to
${\bf P}^1$, one can choose a moving frame ${\bf x}(s),
{\bf y}(s)$ in such way that
$\phi(s)$ is represented by ${\bf x}(s)$. This map can be seen as a
curve in ${\bf P}^1$. Two mappings $\phi$ and $\psi$ are said to be
equivalent if
$\psi=\pi\circ\phi$, with $\pi$ a projective transformation on ${\bf P}^1$.

Flanders considered two extreme situations. The first case corresponds to
$b(s)=0$, $\forall s$. In this case $\phi$ is constant. 
Taking the derivative of $\lambda{\bf x}$, for some $\lambda(s)\ne 0$, we have
by (\ref{structureequations}) that 
$(\lambda{\bf x})'=(\lambda'+a\lambda){\bf x}$. 
Choosing $\lambda\propto\exp[{-\int^s_{s_0}dta(t)}]\ne 0$, we have $(\lambda
{\bf x})'=0$, so that $\lambda {\bf x}$ is a constant representative of $\phi$.

The other case is for $b$ never vanishing. There are only two inequivalent
situations. The first one is when $b$ is either complex or positive. It turns
out that it is always possible to choose the following 
``natural moving frame'' for $\phi$ \cite{Flanders}
\beq
{\bf x}'={\bf y},\qquad{\bf y}'=-k{\bf x}.
\label{nmf}
\eeq
In the other case, corresponding to $b$ real and negative, the natural
moving frame for $\phi$ is
\beq
{\bf x}'=-{\bf y},\qquad{\bf y}'=k{\bf x}.
\label{nmf2}
\eeq

A characteristic property of the natural moving frame is that it is determined
up to a sign with $k$ an invariant. Thus, for example, suppose that for a given
$\phi$ there is, besides (\ref{nmf}), the natural moving frame ${\bf x}'_1={
\bf y}_1$, ${\bf y}'_1=-k_1{\bf x}_1$. Since both ${\bf x}$ and ${\bf x}_1$ are
representatives of $\phi$, we have ${\bf x}=\lambda{\bf x}_1$, so that ${\bf y}=
{\bf x}'=\lambda'{\bf x}_1+\lambda{\bf y}_1$ and $1=[{\bf x},{\bf y}]=\lambda^2
$. Therefore, ${\bf x}_1=\pm{\bf x}$, ${\bf y}_1=\pm{\bf y}$ and $k_1=k$
\cite{Flanders}.

Let us now review the derivation of Flanders formula for $k$. Consider
$s\longrightarrow {\bf z}(s)$ to be an affine representative of $\phi$ and let
${\bf x}(s),{\bf y}(s)$ be a natural frame. Then ${\bf z}=\lambda{\bf x}$ where
$\lambda(s)$ is never vanishing. Now note that, since ${\bf z}'=\lambda'{\bf x}
+\lambda{\bf y}$, we have that $\lambda$ can be written in terms of the area
function $[{\bf z},{\bf z}']=\lambda^2$. Computing the relevant area functions,
one can check that $k$ has the following expression
\beq
2k={[{\bf z},{\bf z}''']+3[{\bf z}',{\bf z}'']\over[{\bf z},{\bf z}']}-
{3\over2}\left({[{\bf z},{\bf z}'']\over[{\bf z},{\bf z}']}\right)^2.
\label{kexpression}
\eeq

Given a function $z(s)$, this can be seen as the non--homogeneous coordinate of
a point in ${\bf P}^1$. Therefore, we can associate to $z$ the map
$\phi$ defined
by $s\longrightarrow (1,z(s))={\bf z}(s)$. In this case we have $[{\bf z},{\bf
z}']=z'$, $[{\bf z},{\bf z}'']=z''$, $[{\bf z},{\bf z}''']=z'''$, $[{\bf z}',
{\bf z}'']=0$, and the curvature becomes \cite{Flanders}
\beq
k={1\over2}\{z,s\}.
\label{kexpression2}
\eeq

For an arbitrary physical state with potential function $W$
we have, 
\beq
W=-{\hbar^2\over4m}\{e^{{2i\over\hbar}S_0},q\}=-{\hbar^2\over2m}k_W,
\label{arbpotW}
\eeq
and similarly for the quantum potential
\beq
Q={\hbar^2\over4m}\{S_0,q\}={\hbar^2\over2m}k_Q,
\label{arbpotQ}
\eeq
where $k_W$ is the curvature associated to the map
\beq
q\longrightarrow(1,e^{{2i\over\hbar}S_0(q)}),
\label{kwmap}
\eeq
while the curvature $k_Q$ is associated to the map
\beq
q\longrightarrow (1, S_0(q)).
\label{kqmap}
\eeq
The function defining the map (\ref{kwmap}) coincides with the
trivializing map, whereas
the Schwarzian identity (\ref{schwarzianidentity})
can be now seen as difference of curvatures
\beq
\left({\partial_q S_0}\right)^2={\hbar^2}k_W-{\hbar^2}k_Q.
\label{SIDascurvatures}
\eeq
The QSHJE (\ref{qshje}) can be written in the form
\beq
{1\over2m}\left({\partial S_0(q)\over\partial q}\right)^2+ W(q)
+{\hbar^2\over2m}k_Q=0.
\label{qshjewc}
\eeq
We therefore note that in the one dimensional QSHJE the
quantum potential is interpreted as a curvature and is an
intrinsically quantum characteristic of the particle.
In higher dimension the curvature term takes the form
\beq
Q=-{\hbar^2\over2m}{\Delta R\over R}
\label{qpforthehumpthtime}
\eeq
and the corresponding form in the relativistic case.
The continuity condition,  eq. (\ref{conteq}), implies
that
\beq
R^2{\partial_i} S_0 = \epsilon_i^{i_2\dots i_D}\partial_{i_2}F_{i_3\dots i_D}, 
\label{dminusform}
\eeq
where $F$ is a $(D-2)$ form. The the $3D$ case $R^2\partial_i S_0$ is the curl 
of a vector, denoted by $B$,
\beq
R^2\nabla S_0 = \nabla\times B. 
\label{R2S0}
\eeq
Hence, the QSHJE, eq. (\ref{qshje}), takes the form 
\beq
(\nabla\times B)^2 = \hbar^2R^3\Delta R+2mER^4. 
\label{Bform}
\eeq
In the time--dependent relativistic case $F$ is a $(D-1)$--form. We have 
\beq 
R^2(\partial_\mu S-eA_\mu)=\epsilon_\mu^{\;\,\sigma_1\ldots\sigma_D} 
\partial_{\sigma_1}F_{\sigma_2\ldots\sigma_D}=\partial^\nu B_{\mu\nu}, 
\label{rdminusform}
\eeq 
that is 
\beq 
R^2={(\partial^\mu S-eA^\mu)\over(\partial S-eA)^2}\partial^\nu B_{\mu\nu}.
\label{RR2S0}
\eeq 
In terms of $B$ and $R$ the RQHJE, which correspond to the real part of eq.
(\ref{gaugepotID}) with $\alpha=i/\hbar$, takes the form
\beq 
\partial^\nu B_{\mu\nu}\partial_\sigma B^{\mu\sigma}+R^4m^2c^2 
-\hbar^2R^3\Box R=0. 
\label{RBform}
\eeq 
Thus, in the higher dimensional case the quantum potential correspond to the 
curvature of the function $R(q)$. We can investigate the properties of the curvature
term by studying the free one dimensional particle, with $W^0(q^0)\equiv0$. In this 
case the Schr\"odinger equation takes the form $${\partial^2\over{\partial q}^2}\psi=0,$$
with the two linearly independent solutions being $\psi^D=q^0$ and $\psi=const$. As
discussed above an essential tenant of the equivalence postulate formalism
is that both solutions of the Schr\"odinger equation must be retained. The linear 
combination of $\psi$ and $\psi^D$, $\psi^D+i\ell\psi$, appears in the solution for
$S_0$, eq. (\ref{ei2s0overbeta}). In the case of the state $W^0(q^0)\equiv 0$, with 
$\psi^D=q^0$ and $\psi=1$ this combination reads $q^0+i\ell_0$, and $\ell_0\equiv\ell$
should have the dimension  of length. On the other hand, from the 
fact that by eq. (\ref{qtoq0}) the trivialising coordinate $q^0$ is given 
by the ratio $\psi^D/\psi$ up to a Mobi\"us transformation, it follows that
$\ell$ and $w=\psi^D/\psi$ have the dimension of length for any state. The 
reduced action $S_0^0$ corresponding to the trivial state $W^0$ is
\beq
{\rm e}^{{2i\over\hbar}S_0^0}={\rm e}^{i\alpha}
{{q^0+i{\bar\ell}_0}\over{q^0-i\ell_0}}, 
\label{s00}
\eeq
and the conjugate momentum $p_0=\partial_{q^0} S_0^0$ has the form
\beq
p_0=\pm{{\hbar (\ell_0+{\bar\ell}_0) }\over {2\vert q^0- i\ell_0\vert^2}}.
\label{p00}
\eeq
It follows that $p_0$ vanishes only for $q^0\rightarrow\pm \infty$,
and is maximised at $q^0=-{\rm Im}\ell_0$
\beq
\vert p_0(-{\rm Im}\,\ell_0)\vert ={\hbar\over{\rm Re}\,\ell_0}.
\label{maxp0}
\eeq
Since ${\rm Re}\,\ell_0\ne0$, $p_0$ is always finite, and $\ell_0$ provides
an ultraviolet cutoff. This property extends to any state \cite{fm}.
It follows that the equivalence postulate implies an ultraviolet 
cutoff on the conjugate momentum. We can contemplate the 
potential form of this ultraviolet cutoff by studying the classical 
limit, 
\beq
\lim_{\hbar\rightarrow0}p_0=0. 
\label{p0hbarto0}
\eeq
From eq. (\ref{p00}) it follows that ${\rm Im}\ell_0$ can be absorbed by a shift
of $q^0$. Hence, in the limit eq. $(\ref{p0hbarto0})$ we can set 
${\rm Im}~\ell_0=0$ and distinguish the cases $q^0=0$ and $q^0\ne0$. 
Let us define $\gamma$ by 
\beq
{\rm Re}\,\ell_0{}_{\stackrel{\sim}{\hbar\longrightarrow0}}\hbar^{\gamma}. 
\label{definegamma}
\eeq
then 
\beq
p_0{}_{\stackrel{\sim}{\hbar\longrightarrow0}}\left\{\begin{array}{ll}
\hbar^{\gamma+1}, & q_0\ne0,\\ \hbar^{1-\gamma},& q_0=0,\end{array}\right. 
\label{p0to0intermsofgamma}
\eeq
and by (\ref{p0hbarto0}) 
\beq
-1<\gamma<1. 
\label{gammagtm1stp1}
\eeq
A constant length having powers of $\hbar$ can be constructed by means of 
the Compton length,
$\lambda_c= \hbar/(mc)$,  and the Planck length, 
$\lambda_p= \sqrt{\hbar G/c^3}$. 
It is also noted that a constant length which is 
independent of $\hbar$ is provided by $\lambda_e=e^2/mc^2$ where $e$ is the electric 
charge. Thus $\ell_0$ can be considered as a suitable function of 
$\lambda_c$, $\lambda_p$ and $\lambda_e$ satisfying the constraint 
(\ref{gammagtm1stp1}). Of the three constants it is noted that $\lambda_p$ 
satisfies the condition  (\ref{gammagtm1stp1}), whereas $\lambda_c$ and 
$\lambda_e$ do not. Therefore, we can set
\beq
{\rm Re}\,\ell_0=\lambda_p= \sqrt{{\hbar G}\over c^3}.
\label{settingell0}
\eeq
With this choice of ${\rm Re}\,\ell_0$, by eq. (\ref{maxp0}), the maximum 
of $\vert p_0\vert$ is given by
\beq
\vert p_0(-{\rm Im}\,\ell_0)\vert =\sqrt{{\hbar c^3}\over G}.
\label{maxp0forell0fixed}
\eeq
The quantum potential associated with the state $W^0$ is given by
\beq
Q^0={\hbar^2\over4m}\{S_0^0,q^0\}= - {{\hbar^2 ({\rm Re}\,\ell_0)^2}\over2m}
{1\over{\vert q^0-i\ell_0\vert^4}}.
\label{Q0pot}
\eeq
We can make naive estimates of this potential. Taking $m\sim 100GeV$; 
${\rm Re}\,\ell_0=\lambda_p\approx10^{-35}m$; and $q^0$ as the size of the 
observable universe $q^0\sim 93 Ly$, gives $\vert Q\vert\sim 10^{-202} eV$. 
A tiny amount of energy indeed! It is noted that the sign of $Q$ is negative
in the one dimensional case. However, there is no reason to expect that this
will be the case in higher dimensions as the curvature of the function $R(q)$ 
is, in principle undetermined. It should be emphasized that this is a naive
analysis, reflecting the appearance of a length scale in the formalism
associated with the intrinsic properties of particle and providing an
internal energy component similar to its mass. In this estimate I have
taken the particle mass to be $100GeV$ and $q$ to be given by the size 
of the observable universe, and the quantum potential is that corresponding to  
the state $W^0$. Taking $q\sim 1m$ yields $\vert Q\vert\sim10^{-96} eV$. 
In this respect a relevant question is what is the effective $q$ that one should 
take. A consistency condition of the equivalence postulate formalism
dictates that the trivializing transformation is continuous on the extended
real line \cite{fm}, {\it i.e.} at $q=\pm\infty$. The question therefore
is what is the effective $q=\pm\infty$ from the elementary particle 
perspective. A more detailed analysis can take in account relativistic 
particles by considering the relativistic quantum Hamilton--Jacobi equation,
and by considering particles with constant, but finite, $W$ functions.  
It is noted that several authors considered the possibility of interpreting the
quantum potential in Bohmian mechanics as the source of dark energy
\cite{gonzalezdiaz}, as well as detailed cosmological scenarios \cite{dcs}.

\section{The multi--particle case}

So far the single particle case was discussed. It is interesting to
study the question of the multi--particle case and whether the 
quantum potential is additive in this case. Furthermore, while the 
single particle case can reveal basic properties of the mathematical 
formalism, the physical case involves the multi--particle state, as
it requires at least an observer and an observee.
In the case of two free particles of energy $E$ and
masses $m_1$ and $m_2$, the QSHJE reads
\beq 
{1\over 2m_1}(\nabla_1S_0)^2+
{1\over 2m_2}(\nabla_2 S_0)^2-E-
{\hbar^2\over 2m_1}{\Delta_1 R\over R}-
{\hbar^2\over 2m_2}{\Delta_2 R\over R}=0. 
\label{twoparticleqshje}
\eeq
The continuity equation is 
\beq 
{1\over m_1}\nabla_1\cdot(R^2\nabla_1 S_0)+ 
{1\over m_2}\nabla_2\cdot(R^2\nabla_2 S_0)=0. 
\label{twoparticlecontcond}
\eeq
Next, we set 
\beq r=r_1-r_2,\qquad r_{c.m.}={m_1r_1+m_2r_2\over
m_1+m_2},\qquad m={m_1m_2 \over m_1+m_2}, 
\label{comcor}
\eeq 
where $r_1$ and $r_2$ are the vectors of the two particles. With respect
to the new variables the equations (\ref{twoparticleqshje}) and
(\ref{twoparticlecontcond}) have the form
\beq 
{1\over 2(m_1+m_2)}(\nabla_{r_{c.m.}} S_0)^2+{1\over 2m}
(\nabla S_0)^2-E-{\hbar^2\over 2(m_1+m_2)}{\Delta_{r_{c.m.}}
R\over R}-{\hbar^2\over 2m}{\Delta R\over R}=0,
\label{tpqshjeinnewvar}
\eeq
\beq 
{1\over m_1+m_2}\nabla_{
r_{c.m.}}\cdot(R^2\nabla_{r_{c.m.}} S_0)+
{1\over m}\nabla\cdot(R^2\nabla S_0)=0, 
\label{tpcontconinnewvar}
\eeq where
$\nabla$ ($\nabla_{r_{c.m.}}$) and $\Delta$ ($\Delta_{r_{c.m.}}$)
are the gradient and Laplacian with respect to the components of
the vector $r$ ($r_{c.m.}$). These equations can be decomposed
into the equations for the center of mass $r_{c.m.}$ and those for
the relative motion. Ref. \cite{matone} focussed on the equations
for the relative motion and argued that the quantum potential
is at the origin of the gravitational interactions. Estimating the 
energy due to gravitational interaction between two protons
at a distance of $1fm=10^{-15}m$ gives $\vert E\vert\sim10^{-31}eV$, 
whereas estimating it from the quantum potential yields 
$\vert E\vert\sim 10^{-33}eV$, which is not too far off. 
While this is an intriguing proposition, it seems that the
constructive approach to incorporate gravity into the 
formalism is to extend the cocycle and quadratic identities 
to curved backgrounds, which is left for future work. Instead I focus 
here on the equations for motion of the centre of mass, which take the form 
\beq 
{1\over 2(m_1+m_2)}(\nabla_{r_{c.m.}} S_0)^2-{\tilde E}-
{\hbar^2\over 2(m_1+m_2)}{\Delta_{r_{c.m.}}
R\over R}=0,
\label{comhje}
\eeq
\beq 
{1\over(m_1+m_2)}\nabla_{
r_{c.m.}}\cdot(R^2\nabla_{r_{c.m.}} S_0)=0. 
\label{comce}
\eeq 
The main point to note is that the effective mass in the relative
motion is the reduced mass, whereas the effective mass in the centre 
of mass motion is additive, which is what we expect physically. Hence,
for large number of particles the mass and potential energies are 
summed. In the equivalence postulate framework the mass and 
quantum energy of elementary particles are correlated, and are different 
attributes of the same particle. Finally, the same arguments could 
have been presented by using the relativistic form of the equations. 
However, non--relativistic treatment is viable given that the majority of
matter in the universe is expected to be non--relativistic. 

\section{Conclusions}

Contemporary observations at the smallest and highest length scales
are well accounted for by the Standard Model of particle physics and 
by general relativity, respectively. Yet the synthesis of these two basic
theories remains elusive. 
By hypothesizing that all physical systems labelled by a
potential function $W(q)$ are connected by coordinate
transformations, 
the equivalence postulate of quantum mechanics
offers an axiomatic starting point for formulating quantum field theories 
and quantum gravity.
A somewhat heuristic view of the 
equivalence postulate procedure is the following.
In the classical Hamilton--Jacobi formalism one seeks 
a transformation from an a non--trivial Hamiltonian
to a trivial Hamiltonian. 
The transformations are canonical and treat the phase
space variables as independent variables. This ensures 
that the classical equations of motion are preserved. 
It ensures that the classical path is selected. 
The key property of quantum mechanics is that the 
phase space variables are not independent,
but satisfy the basic quantum mechanical relation
\beq
\left[{\hat q},{\hat p}\right]=i\hbar~.
\label{basicqm}
\eeq
Therefore, we can view the equivalence postulate procedure
as implementing the Hamilton--Jacobi trivialisation 
algorithm, but taking the phase space variables to be
related by the generating function $S(q)$, via the 
relation (\ref{peqpsdq}). The solution to this
problem is not given by the CSHJE, eq. (\ref{cshje}) but rather 
by the QSHJE, eq. (\ref{qshje}). Covariance of the QSHJE
then implies that all physical systems labelled by 
the potential function $W(q)$ are connected by the
coordinate transformations. This view is in line 
with the picture that quantum mechanically 
all possible paths are allowed, and we 
may imagine that different path correspond to 
different $W(q)$ labels. The equivalence postulate 
formalism is encoded in two basic identities. 
The first is the cocycle condition, eq. (\ref{cocycle}), 
which embodies the symmetry properties underlying 
quantum mechanics in this formalism. The second is the
quadratic identity, eq. (\ref{schwarzianidentity}), which 
is the manifestation of the Quantum Hamilton Jacobi 
Equation and is compatible with the equivalence hypothesis. 
Both of these key ingredients admit suitable generalisations 
in higher dimensions with Euclidean or Minkowski metrics.
A key role in this approach is played by the various dualities 
\cite{fm,fm1}, which are implemented as Legendre transformations. 
These Legendre dualities represent a complementary facet of the 
underlying physical principles. In this respect they show that 
the different physics variables used to characterise the physical systems
are related by the dualities and none is superior to the other.

In view of the observational supremacy of the Standard Model and General
relativity in their respective domains, we may question how the equivalence 
postulate formalism is manifested observationally. Given that
quantum gravity effects are likely to be notoriously small, we may
expect them to generate deviations in the most extreme regimes
of contemporary observations.  
Indeed, recently the OPERA collaboration announced intriguing results on the
superluminal propagation of neutrinos \cite{opera}. While the fate of
the results of this experiment is in doubt \cite{icarus}, 
it is clear that the neutrino sector will provide  
experimental data that will continue to probe the validity of the Standard Model. 
In the context of the equivalence postulate formalism
it was shown in refs. \cite{marco, odatepoqm, floyd2}
that the quantum potential in this formalism leads to 
dispersion relations that modify the classical relativistic 
energy--momentum relation. Such modifications
of the relativistic energy--momentum relation are expected
in different approaches to quantum gravity and the
relevant question is whether they are sufficiently large to be observable. 
In this paper I proposed that the quantum potential that arises
in the equivalence postulate formulation can be regarded as
an intrinsic curvature term, which characterises elementary particles, 
and may be interpreted as dark energy. 

\section*{Acknowledgements}

I would like to thank the theoretical physics department at
Oxford University for hospitality. 
This work was supported in part by the STFC (PP/D000416/1).

\end{document}